\documentclass[aps,prb,twocolumn,amsmath,floatfix]{revtex4-2}
\usepackage{dcolumn}    
\usepackage{bm}         
\usepackage{hyperref}   
\usepackage[usenames,dvipsnames]{xcolor}
\usepackage[normalem]{ulem} 
\usepackage{graphicx}
\usepackage[english]{babel}

\usepackage[utf8]{inputenc}
\usepackage[T1]{fontenc}

\usepackage{scalerel,stackengine}
\stackMath
\newcommand\reallywidehat[1]{%
\savestack{\tmpbox}{\stretchto{%
  \scaleto{%
    \scalerel*[\widthof{\ensuremath{#1}}]{\kern-.6pt\bigwedge\kern-.6pt}%
    {\rule[-\textheight/2]{1ex}{\textheight}}
  }{\textheight}%
}{0.5ex}}%
\stackon[1pt]{#1}{\tmpbox}%
}

\begin{document}

\title{Comment on ``Theoretical analysis of quantum turbulence using the Onsager ideal turbulence theory''}

\author{Giorgio Krstulovic}
\affiliation{%
  Université Côte d'Azur, Observatoire de la Côte d'Azur, CNRS,
  Laboratoire Lagrange,  Boulevard de l'Observatoire CS 34229 - F 06304 NICE Cedex 4, France.
}
\author{Victor L'vov}
\affiliation{Dept. of Chemical and Biological Physics, Weizmann Institute of Science, Rehovot 76100, Israel.}%
\author{Sergey Nazarenko}
\affiliation{Université Côte d'Azur, CNRS, Institut de Physique de Nice. Avenue Joseph Vallot 06108 Nice, France}

\date{\today}


\begin{abstract}
        In a recent paper [T. Tanogami Phys. Rev. E 103, 023106 ] proposes a scenario for quantum turbulence where the energy spectrum at scales smaller than the inter-vortex distance is dominated by a \emph{quantum stress cascade}, in opposition to Kelvin wave cascade predictions. The purpose of the present comment is to highlight some physical issues in the derivation of the \emph{quantum stress cascade}, in particular to stress that quantization of circulation has been ignored. 
\end{abstract}

\maketitle

In a recent paper, T.~Tanogami presents a theoretical investigation of quantum turbulence at very low temperatures by adapting standard techniques used in classical hydrodynamics \cite{Tanogami_TheoreticalAnalysisQuantum_2021}.  Following Onsager's ideas of classical turbulence \cite{Eyink_OnsagerTheoryHydrodynamic_2006}, Tanogami proposes a double energy cascade scenario where the energy spectrum $E^v(k)$ behaves as
\begin{equation}
 E^v(k)\sim \begin{cases}
               C_{\rm large}\,k^{-5/3} &\text{for } \ell_{\rm large}^{-1}\ll k \ll \lambda^{-1} \\
               C_{\rm small}\,k^{-3}   &\text{for } \lambda^{-1} \ll k \ll \ell_{\rm small}^{-1}.
         \end{cases},\label{Eq:DoubleCascade}
\end{equation}
Here $k$ is the wave vector, and $C_{\rm large}$ and $C_{\rm small}$ are positive constants. $\ell_{\rm large}$ is a scale that can be identified with the inertial scale of turbulence and $\ell_{\rm small}$ is defined using the quantum stress cospectrum (see  \cite{Tanogami_TheoreticalAnalysisQuantum_2021}). Then, Tanogami defines the \emph{quantum baropycnal work} flux $\Lambda_\ell^{\Sigma}$ and identifies the scale $\lambda$ as the scale at which $\Lambda_\ell^{\Sigma}$ becomes constant. Finally, he associates $\lambda$ with the mean inter-vortex distance $\ell_i$.

It has been largely accepted by the community that at scales smaller than the inter-vortex distance, the energy spectrum should display an energy cascade where the physics is dictated by the Kelvin wave cascade \cite{Kozik_KelvinWaveCascadeDecay_2004,L'vov_WeakTurbulenceKelvin_2010,Laurie_InteractionKelvinWaves_2010}. With his derivation, Tanagomi questions this cascade and proposes a completely different mechanism. In this comment we argue that the $k^{-3}$ \emph{quantum stress cascade} is based on unphysical assumptions, and therefore such a scenario can not take place. Basically, Tanagomi's derivation completely ignores the quantization of velocity circulation, a crucial property of quantum turbulence that can not be neglected at scales smaller than $\ell_i$. We describe in the following our main criticisms on the double cascade scenario proposed in \cite{Tanogami_TheoreticalAnalysisQuantum_2021}.

The main issue of Tanagomi's work is the starting equations used to apply and adapt to the quantum case, the standard techniques used in compressible classical turbulence. It is explicitly stated, already in the abstract, that three-dimensional quantum turbulence is investigated by using the Gross-Pitaevskii (GP) equation, but this statement is highly misleading. The GP equation is indeed an excellent theoretical framework to study low-temperature quantum turbulence because quantum vortices naturally arise as topological defects of the macroscopic wave function. As a consequence, the velocity circulation around a quantum vortex is quantized. However, the actual starting equations of Tanagomi's derivation are the dispersive Euler equations, obtained after using the Madelung transformation in the GP equation. This transformation writes the complex wave function $\psi$ as 
\begin{equation}
  \psi({\bf x})=\sqrt{\rho({\bf x})}e^{i\theta({\bf x})},\label{eq:madelung}
\end{equation}
where $\rho$ can be identified with the fluid density and the phase $\theta$ defines the velocity field through the relation ${\bf v}= (\hbar/m)\nabla\theta$, with $\hbar$ the reduced Planck constant and $m$ the boson mass. Indeed, after introducing \eqref{eq:madelung} into the GP equation, one obtains the continuity equation for the density (Eq.(2) of Tanogami’s work) and a modified Bernouilli equation for the phase $\theta$. Only after taking the gradient of the Bernouilli equation and rearranging terms one obtains the momentum equation for the velocity (Eq.(3) of Tanogami’s work). 

At the vortex core, $\psi$ vanishes and therefore its phase is not defined. As a consequence, the dispersive Euler equation is mathematically equivalent to the GP equation only in absence of vortices, and therefore the Kolmogorov-Onsager ideas of turbulence can not be there directly applied. 
The dispersive Euler equations are however useful to provide a phenomenological hydrodynamic interpretation of the GP equation at scales where the quantization of circulation can be ignored, that is at scales much larger than the inter-vortex distance $\ell_i$. In order to give some meaning to dispersive Euler equation at scales smaller than $\ell_i$, such equations have to be supplied with information about the location of quantum vortices, that reduces in these variables to extremely complex moving boundary conditions on three-dimensional curves where density vanishes and circulation is quantized. Furthermore, even if $\rho$ and ${\bf v}$ are such that at $t=0$ they represent a quantum turbulent state, the quantization of circulation will not be preserved by the dispersive Euler dynamics for $t>0$. For the previous reasons, Tanagomi's work could provide, in principle, a rigorous derivation of the energy spectrum based on the Onsager conjecture and a phenomenological model of quantum turbulence, only at scales $k\ll \ell_i^{-1}$, i.e at scales where quantum turbulence displays a classical behavior. Therefore the \emph{quantum stress cascade} has no physical relevance and should not be considered as a possible scenario.

In more specific terms, in the main text Tanagomi's derivation assumes some regularity of the velocity field that does not hold for quantum vortices. It is assumed that the velocity field is H{\"o}lder continuous with exponent $h\in(0,1]$, i.e.
\begin{equation}
  \delta{\bf v}({\bf r},{\bf x})={\bf v}({\bf r}+{\bf x})-{\bf v}({\bf x})=O(|{\bf r}|^h),
\end{equation} 
for $\ell=|{\bf r}|\to0$. Note that a quantum vortex corresponds to a regularity of $h=-1$. In order to overcome this issue, the author considers a domain $\Omega$ that excludes any possible point ${\bf x}$ having a local H{\"o}lder exponent with $h<0$. Doing so, all vortices are excluded from the domain. Therefore, only far field contributions of the velocity field are retained. Even if the notion of quantized vortices were somehow introduced in the dynamics of the dispersive Euler equation, excluding the energy of this domain would certainly miss most contributions arising from Kelvin waves. To further avoid this possible issue, the author devotes one Appendix to generalize the calculations using Besov spaces. In this approach, only the $L^p$ norm of the velocity increments is assumed to have such restricted regularity, allowing in principle the velocity field to have a local negative H{\"o}lder exponent. Such an assumption is far from being justified physically. Note that, for a quantum vortex, the velocity diverges but the density simultaneously vanishes at its position, in such a way that the wave function, and the momentum and energy densities are completely regular fields. This singularity is only a trivial consequence of the Madelung transformation not being defined at the vortex core.

In addition, we would like to remark that the \emph{quantum-stress cascade} predicted by Tanagomi results from the quantum pressure term of the dispersive Euler equation. Such term, is independent of the GP non-linearity and therefore, the same analysis could be applied to the (linear) Schrodinger equation. In this equation, being linear, we can not expect an energy transfer along scales. Perhaps, the $k^{-3}$ is a consequence of the regularity of the wave function at small scales, in agreement with Tanagomi's choice of a local H{\"o}lder exponent $h=1$ in this range. Such a scaling could emerge at scales $\ell\ll\xi$, however, as mentioned in his work, it could be much shallower due to nonlinearity depletion.  

Finally, from a practical point of view, if such a \emph{quantum-stress cascade} displaying a $k^{-3}$ scaling exits at scales between $\ell_i$ and the healing length $\xi$, it should be overwhelmed by Kelvin waves (KW), which display a much less steep spectra \cite{Kozik_KelvinWaveCascadeDecay_2004,L'vov_WeakTurbulenceKelvin_2010,Laurie_InteractionKelvinWaves_2010}. Also, we would like to remark that there exist several works using the GP equation where the KW cascade has been observed, which we summarize in the following. In \cite{Krstulovic_KelvinwaveCascadeDissipation_2012}, the KW cascade was observed by directly tracking perturbed straight vortex lines. In that setting, there was not Kolmogorov cascade as by construction $\ell_i\sim\ell_{\rm large}$, but the KW cascade was observed to be compatible with weak wave turbulence predictions. Later, in \cite{ClarkdiLeoni_SpatiotemporalDetectionKelvin_2015}, KWs were studied in turbulent tangles (exhibiting a Kolmogorov spectrum) by using spatio-temporal filtering. Then, by tracking large vortex rings of turbulent quantum tangles, Villois et al. \cite{Villois_EvolutionSuperfluidVortex_2016} observed the development of a KW cascade with a spectrum supporting L'vov \& Nazarenko prediction \cite{L'vov_WeakTurbulenceKelvin_2010} for more than one decade. Finally, several simulations, using resolutions up to $4096^3$ collocations points, have observed a Kolmogorov scaling range, followed by a KW cascade at scales smaller than $\ell_i$ \cite{ClarkdiLeoni_DualCascadeDissipation_2017,Shukla_QuantitativeEstimationEffective_2019,Muller_KolmogorovKelvinWave_2020a}. In particular, in M{\"u}ller \& Krstulovic \cite{Muller_KolmogorovKelvinWave_2020a}, the  L'vov \& Nazarenko prediction was observed, including the scaling with the energy flux. In summary, we believe that there is enough evidence supporting the scenario where in quantum turbulence the Kolmogorov energy cascade is followed by the KW energy cascade.


\begin{acknowledgments}
  We acknowledge useful scientific discussions with M.~Brachet and N.~P. M{\"uller}. GK was supported by the Agence Nationale de la Recherche through the project GIANTE ANR-18-CE30-0020-01. GK and SN were also supported by the Simons Foundation Collaboration grant Wave Turbulence (Award ID 651471).
  \end{acknowledgments}



%

\end{document}